\documentclass[manuscript]{aastex}

\shorttitle{Nova Scorpii 1941 (V697 Sco): A Probable Intermediate Polar}
\shortauthors{Brian Warner \& Patrick A. Woudt}

\begin{document}

\title{Nova Scorpii 1941 (V697 Sco): A Probable Intermediate Polar\altaffilmark{1}}

\author{Brian Warner and Patrick A. Woudt}
\affil{Department of Astronomy, University of Cape Town, Rondebosch, 7700
South Africa}
\email{warner@physci.uct.ac.za, pwoudt@circinus.ast.uct.ac.za}

\altaffiltext{1}{This paper uses observations made from the South African Astronomical Observatory (SAAO).}

\begin{abstract}
V697 Sco, the remnant of Nova Scorpii 1941 and currently at $V \sim 20.0$, is found from photometric
observations to have the characteristics of an intermediate polar (IP) with an orbital period ($P_{orb}$) of 4.49 h and
a rotation period ($P_{rot}$) of 3.31 h. It therefore appears to be a member of the rare class of IPs where
$P_{rot} \sim P_{orb}$, which are probably discless systems. The prominence of the modulation at
${{1}\over{2}}$ $P_{rot}$, and its orbital sidebands, indicates two-pole accretion.
\end{abstract}

\keywords{cataclysmic variables, stars: binaries: close, star: individual (V697 Sco)}

\section{INTRODUCTION}

   Nova Scorpii 1941 (V697 Sco) was discovered by Margaret Mayall on objective prism 
plates at a photographic magnitude of 10.2 (Mayall 1947). The spectrum was of a nova 
at a late stage of development and it was later deduced that at its brightest it had 
reached $V \sim 8$ (Payne-Gaposchkin 1957). Duerbeck (1981) assigns the light curve 
to his type A, which has a smooth fast decline without disturbances, and a very fast 
speed class with $t_3$ $<$ 15 d. The Downes et al. (1997) catalogue gives a magnitude at minimum 
of 17, measured on a J plate. We find that currently it is about 3 mag fainter than that.
The amplitude of eruption is compatible with the amplitude -- $t_2$ relationship (e.g., 
Warner 1995).

\section{PHOTOMETRIC OBSERVATIONS}

Our high speed photometry of V697 Sco was carried out on the 74-in reflector at the
Sutherland site of the South African Astronomical Observatory in 2001 and 2002. We used 
the University of Cape Town's CCD photometer (O'Donoghue 1995) in frame transfer mode 
and without any filter. Photometric calibration was achieved by observation of hot 
white dwarf standards. The finding chart in the Downes et al. catalogue shows
that V697 Sco is an isolated star, but crowding among the nearby reference stars restricted 
observations to conditions of particularly good seeing. 
Our observational log is given in Table~\ref{tab1}.

\begin{figure}
\plotone{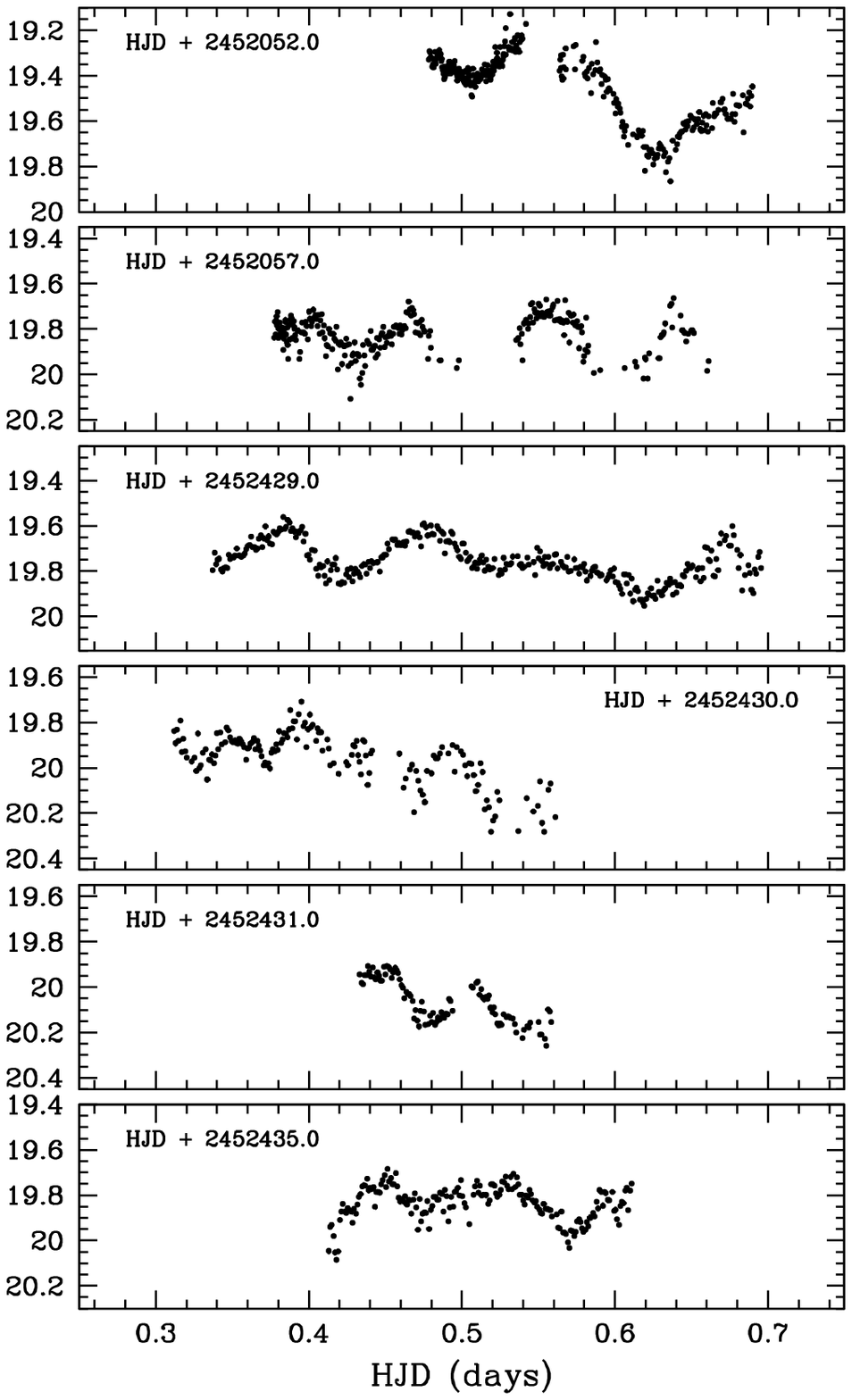}{}
\caption{The light curves of V697 Sco, obtained in May 2001 (the top two light curves) and June 2002
(the lower four light curves).}
\label{v697scolc}
\end{figure}

The light curves of V697 Sco are displayed in Figure~\ref{v697scolc} and show V697 Sco
to have a complicated waveform, which initially we found difficult to understand.
Fourier transforms (FTs) of these light curves reveal that there is an underlying 
simplicity: an orbital modulation (non-sinusoidal) and a spin period with orbital
sidebands, characteristic of an intermediate polar (IP).

\section{ANALYSIS OF THE LIGHT CURVES}

\begin{figure*}
\plotone{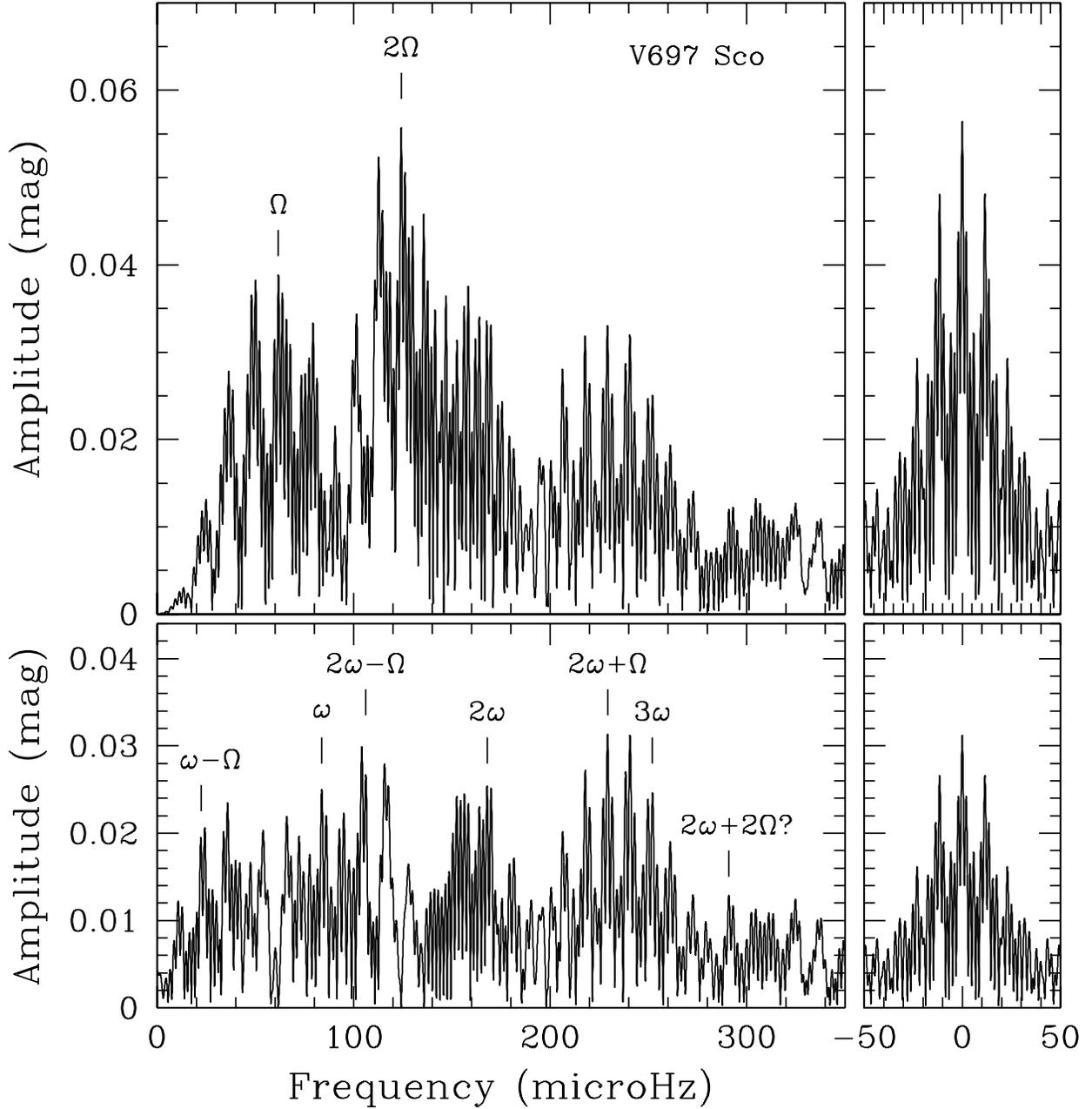}
\caption{The Fourier transform of the June 2002 observations of V697 Sco. The top left panel shows
all the data; the orbital frequency and its first harmonic are indicated. The top right panel shows
the window function, scaled to the highest peak in the top left panel.
The bottom left panel illustrates the same data, but now prewhitened at the orbital frequency and
its first harmonic.  The identified suite of frequencies is marked; they are also listed in
Table 2. Again the window function (scaled to the highest peak) is shown in the bottom right panel.}
\label{v697scoft}
\end{figure*}

The FT of the total data set obtained in June 2002
(with the means, and linear trends subtracted for each night)
is shown in Figure~\ref{v697scoft}. The
spectral window shows that, if the data were noise-free, reasonably unambiguous results should
be obtained. A number of window patterns are easily identifiable in the FT, the lowest frequencies
of which are at 61.8 $\mu$Hz and its first harmonic. Despite the presence of 1 d$^{-1}$ and
6 d$^{-1}$ aliases (resulting from the distribution of observations: Table~\ref{tab1}), the
choice of these frequencies is unambiguous if the requirement that they are harmonically related
is enforced (see Table~\ref{tab2}). We identify the lowest frequency as the orbital frequency
$\Omega = 2 \pi / P_{orb}$. This gives $P_{orb}$ = 4.49 h.

To reveal other frequencies more clearly, we prewhitened the light curves at $\Omega$ and $2 \Omega$,
producing the FT shown in the lower panel of Figure~\ref{v697scoft}. Despite the apparent complexity
of the FT, it consists largely of a number of harmonics of the frequency $\omega$ = 83.8 $\mu$Hz, their
differences with $\Omega$ and $2 \Omega$, and the associated window patterns. This is shown
in Table~\ref{tab2} where the `predicted' frequencies of the components $n \omega \pm m \Omega$ (adopting
the values of $\omega$ and $\Omega$ given above) are compared with observed features in the FT.
The suggested identifications have been marked in Figure~\ref{v697scoft}. The modulation
at $2 \omega + 2 \Omega$ is marked with a query because although there is a spike at the `predicted' 
value, it is only at the level of the noise. There is no modulation present at the predicted 
value of $2 \omega - 2 \Omega$.

The frequencies displayed by V697 Sco indicate that it is an asynchronous magnetic rotator, i.e. an
Intermediate Polar (IP) with an orbital period of 4.49 h and a white dwarf rotation period 
$P_{rot} = 2 \pi / \omega = 3.31$ h.

The unusual appearance of the light curve of V697 Sco (i.e., compared to other IPs) is due
to the presence of several modulations having similar amplitudes, with resulting beats between them.

\section{DISCUSSION}

V697 Sco is unique in having $P_{rot} \sim P_{orb}$ at such a large value of $P_{orb}$: there are
two other such IPs, EX Hya (e.g., Warner 1995) and V1025 Cen (Hellier, Wynn \& Buckley 2002), but these
are both short period systems. The properties of the three systems are listed in Table~\ref{tab3}.

The suite of modulations seen in V697 Sco is quite different from the IP HZ Pup (Nova Pup 1963), 
where the more conventional $\omega \pm n \Omega$ is observed (Abbott \& Shafter 1997). The strength
of the $2 \omega$, and the $2 \omega - \Omega$ and $2 \omega + \Omega$ components (which are the $2 \omega$
amplitude modulated at the orbital frequency) in V697 Sco suggests two-pole accretion onto the
primary; the appearance of $\omega$ and $\omega - \Omega$ components indicates that the accretion
rates are different for the two poles.

Almost all IPs have $P_{rot} \sim 0.1 P_{orb}$, but King \& Wynn (1999) have drawn attention
to the possibility of spin equilibrium near $P_{rot} \sim P_{orb}$. They noticed that in EX Hya
and V1025 Cen the corotation radius $r_{co}$ (in a Keplerian disc) of the white dwarf primary is
similar to the distance $R_{L_1}$ from the centre of the white dwarf to the inner Lagrangian point 
at $L_1$. Table~\ref{tab3} illustrates this for the three IPs: we have adopted 0.6 M$_{\odot}$ for 
the primaries of EX Hya and V1025 Cen, but 1.0 M$_{\odot}$ for the primary of V697 Sco, which is a
nova remnant (which characteristically have higher masses because of the selection effect arising
from more frequent nova eruptions at higher masses). The mass $M_1(2)$ of the secondary is obtained
from equation 2.100 of Warner (1995). In all three stars $r_{co}$ is within $\sim$ 20\% of $R_{L_1}$. In 
effect, this implies that the magnetic field of the primary is strong enough to control gas
flow from close to $L_1$, probably producing a `discless' system. Nevertheless, part of the transferring gas may 
be held in a ring or torus, which may eventually be dumped onto the primary, producing short-lived 
outbursts as seen in EX Hya. Some of the gas, however, is centrifuged out along the field lines,
resulting in equilibrium $P_{rot}$ at a much larger value.

King \& Wynn (1999) conclude that an IP equilibrium with $P_{rot} \sim P_{orb}$ should only
be possible for $P_{orb} \la 2$ h. The requirement for such equilibrium is that the magnetic moment
$\mu (1)$ of the primary should be large enough to produce a field at $L_1$ that can control 
gas flow, but not strong enough to synchronize the system as a polar.
The latter requires $\mu (1) \mu (2) / a^3 < 2 \pi \dot{M} R^2_{L_1} / P_{orb}$ (King, Frank \& 
Whitehurst 1991; King \& Wynn 1999) where $\dot{M}$ is the rate of mass transfer and $a$ is the diameter
of the orbit. It is possible to generalise this condition by using 
$\mu_{33} (2) = 2.8 P_{orb}^{9/4}$(h) for the secondary (Warner 1996; $\mu_{33} = \mu / 10^{33}$ G cm$^3$) 
and equation 2.4c of Warner (1995)
to derive, for {\sl non}-synchronization, 

\begin{equation}
\mu_{34} (1) \la 0.21 M_1^{5/3}(1) \dot{M}_{17}
\end{equation}

where very weak dependences on $P_{orb}$ and $q = M(2)/M(1)$ have been supressed.

For EX Hya and V1025 Cen, for which $\dot{M}_{17} \sim 0.1$ (e.g., Warner 1995), we have
$\mu_{34}(1) \la 0.01$; and for V697 Sco, where we use $\dot{M}_{17} \sim 3$ appropriate to its
longer $P_{orb}$, $\mu_{34} (1) \la 0.6$. IPs typically have $0.006 < \mu_{34}(1) < 0.9$ (Warner 1995)
so the asynchronism of the three stars in Table~\ref{tab3}, even V697 Sco with its long
period, is compatible with theory.

There are two other IPs where the first harmonic $2 \omega$ of the rotation frequency is prominent
in the optical region: YY Dra (Patterson et al.~1992) and RX\,J0558+53 (Allan et al.~1996).
These have $P_{orb}$ = 3.96 h and 4.15 h and $P_{rot}$ = 8.8 min and 9.1 min, respectively, and do not
show strong $2 \omega \pm \Omega$ sidebands. V697 Sco appears to be in a class of its own.

\section*{ACKNOWLEDGEMENTS}

BW is funded by the University of Cape Town. PAW is funded partly by strategic funds
made available to the University of Cape Town to BW and partly by the National
Research Foundation.

\clearpage

\clearpage

\clearpage

\begin{table}
\begin{center}
\caption{Observing log}
\begin{tabular}{llllcc}
\tableline\tableline
Run No. & Date of obs. & HJD of first obs. & Length (h) & $t_{in}$ (s) &   $<$V$>$ (mag) \\
        & (start of night) & (+2452000.0)  &            &               &     \\
\tableline
S6225 & 22 May 2001 &    52.47848   &   5.07       &  30, 60     &  19.5   \\
S6234 & 27 May 2001 &    57.37757   &   6.80       &  60         &  19.8   \\
S6410 & 3 June 2002 &   429.33729   &   8.60       &  90         &  19.8   \\
S6413 & 4 June 2002 &   430.31182   &   5.98       & 120         &  20.0   \\
S6418 & 5 June 2002 &   431.43333   &   3.00       &  90         &  20.1   \\
S6430 & 9 June 2002 &   435.41295   &   4.75       &  90         &  19.8   \\
\tableline
\label{tab1}
\end{tabular}
\end{center}
\end{table}

\clearpage

\begin{table}
\begin{center}
\caption{The suite of frequencies and their aliases. Amplitudes (in magnitudes) are given in brackets. The selected
frequencies are listed in bold.}
\begin{tabular}{crrr}
\tableline\tableline
ID. & Frequency ($\mu$Hz) & Frequency ($\mu$Hz) & `Predicted' ($\mu$Hz)\\ 
\tableline
$\Omega$ &    {\bf    61.8 (0.039)} &   63.8 (0.037) & \\
$2 \Omega$ &  {\bf   124.2 (0.056)} &  126.2 (0.051) & 123.6 \\
$\omega - \Omega$&  {\bf 22.3 (0.020)} & 24.3 (0.021) & 22.0 \\
$\omega$ &    {\bf    83.8 (0.025)} &   86.1 (0.022) &  \\
$2 \omega - \Omega$ & 104.2 (0.030) & {\bf 106.2 (0.027)} & 105.8 \\
$2 \omega$&   {\bf   168.0 (0.025)} & 170.0 (0.025) & 167.6 \\
$2 \omega + \Omega$ & {\bf 229.3 (0.031)} & & 229.4  \\
$3 \omega$&   249.8 (0.024) & {\bf 252.1 (0.025)} & 251.4 \\
$2 \omega + 2 \Omega$ & {\bf 291.0 (0.013)} & 293.1 (0.012) & 291.2 \\
\tableline
\label{tab2}
\end{tabular}
\end{center}
\end{table}

\clearpage

\begin{table}
\begin{center}
\caption{Properties of three Intermediate Polars}
\begin{tabular}{lccccc}
\tableline\tableline
Star & $P_{orb}$ (h) & $P_{rot}$ (h) & $M_1(1)$ & $M_1(2)$ & $r_{co}/R_{L_1}$ \\
     &               &               & (assumed) & (assumed) &                \\
\tableline
V1025 Cen & 1.41 & 0.60 & 0.6 & 0.11 & 0.77 \\
EX Hya    & 1.63 & 1.12 & 0.6 & 0.12 & 1.1  \\
V697 Sco  & 4.49 & 3.31 & 1.0 & 0.42 & 1.2  \\
\tableline
\label{tab3}
\end{tabular}
\end{center}
\end{table}

\end{document}